\documentclass[twocolumn,showpacs,aps,prl,superscriptaddress]{revtex4}

\usepackage{graphicx}
\usepackage{dcolumn}
\usepackage{amsmath}
\usepackage{epsfig}

\RequirePackage{xspace}

\input pubboard/babarsym

\newcommand{\BABARPubYear}    {08}
\newcommand{\BABARPubNumber}  {003}

\newcommand{\SLACPubNumber} {13142}

\def\CPp                {\ensuremath{C\!P\!+}\xspace}
\def\CPm                {\ensuremath{C\!P\!-}\xspace}

\def\btodk    {\ensuremath {B^-{\to}D^0 K^-}}
\def\btdk     {\ensuremath {B^\pm{\to}D^0 K^\pm}}

\def\btodp    {\ensuremath {B^-{\to}D^0 \pi^-}}
\def\btdp     {\ensuremath {B^\pm{\to}D^0 \pi^\pm}}

\def\btodh    {\ensuremath {B^-{\to}D^0 h^-}}

\def\dotopp   {\ensuremath {D^0{\to}\pi^-\pi^+}}

\begin{document}

{\pagestyle{empty}
\begin{flushleft}
\babar-PUB-\BABARPubYear/\BABARPubNumber \\
SLAC-PUB-\SLACPubNumber \\
arXiv:0802.4052v1 \\
\end{flushleft}

\preprint{\babar-PUB-\BABARPubYear/\BABARPubNumber} 
\preprint{SLAC-PUB-\SLACPubNumber}

\title{
  {\large \bf
    Improved measurement of \boldmath{\CP} observables in \boldmath{$B^\pm \to D^0_{\CP} K^\pm$} decays
  }
}

%
\author{B.~Aubert}
\author{M.~Bona}
\author{Y.~Karyotakis}
\author{J.~P.~Lees}
\author{V.~Poireau}
\author{E.~Prencipe}
\author{X.~Prudent}
\author{V.~Tisserand}
\affiliation{Laboratoire de Physique des Particules, IN2P3/CNRS et Universit\'e de Savoie, F-74941 Annecy-Le-Vieux, France }
\author{J.~Garra~Tico}
\author{E.~Grauges}
\affiliation{Universitat de Barcelona, Facultat de Fisica, Departament ECM, E-08028 Barcelona, Spain }
% ...
\collaboration{The \babar\ Collaboration}
\noaffiliation

\date{\today}

\begin{abstract}
We present a study of the decay $B^-{\to}\Dz_{(\CP)}K^-$ and its charge
conjugate, where $\Dz_{(\CP)}$ is reconstructed in both a non-\CP
flavor eigenstate and in \CP (\CP-even and \CP-odd) eigenstates, based
on a sample of 382 million $\Y4S\to\BB$ decays collected with the
\babar\ detector at the PEP-II \epem storage ring. We measure the
direct \CP asymmetries $A_{\CP\pm}$ and the ratios of the branching
fractions $R_{\CP\pm}$:
$A_{\CPp} = 0.27\pm 0.09\stat\pm 0.04\syst$,
$A_{\CPm} =-0.09\pm 0.09\stat\pm 0.02\syst$,
$R_{\CPp} = 1.06\pm 0.10\stat\pm 0.05\syst$,
$R_{\CPm} = 1.03\pm 0.10\stat\pm 0.05\syst$.
We also express the results in terms of the so called Cartesian
coordinates $x_+$, $x_-$, and $r^2$:
$x_+ = -0.09\pm 0.05\stat\pm 0.02\syst$,
$x_- =  0.10\pm 0.05\stat\pm 0.03\syst$,
$r^2 =  0.05\pm 0.07\stat\pm 0.03\syst$.
These results will help to better constrain the phase parameter $\gamma=\arg(-V_{ud}V_{ub}^*/V_{cd}V_{cb}^*)$
of the Cabibbo-Kobayashi-Maskawa quark mixing matrix.
\end{abstract}


\maketitle

The angle $\gamma=\arg(-V_{ud}V_{ub}^*/V_{cd}V_{cb}^*)$ is one of the
least precisely known parameters of the corresponding unitarity triangle of the
Cabibbo-Kobayashi-Maskawa matrix~\cite{ckmmatrix}.
There are many proposals on how to measure $\gamma$ involving charged $B$ decays.
The $B^-{\to}D^{(*)0}K^{(*)-}$ decay mode~\cite{chargeconj}, which exploits the
interference between $b\to c\bar{u}s$ and $b\to u\bar{c}s$ decay
amplitudes, is one of the most important of these~\cite{glw1991,others_btdk}.
In this paper we use a
theoretically clean measurement technique suggested by Gronau, London, and Wyler
(GLW). It exploits the interference between $B^-{\to} \Dz K^-$ and $B^-{\to}
\Dzb K^-$ decay amplitudes, where the \Dz and \Dzb mesons decay to the same \CP
eigenstate~\cite{glw1991}. We express the results in terms of the commonly used ratios $R_{\CP\pm}$ of
charge-averaged partial rates and of the partial-rate charge
asymmetries $A_{\CP\pm}$,
 
\begin{eqnarray}
	\label{eq:Rcp}
	R_{\CP\pm} &=& \frac{\Gamma(B^-{\to}\Dz_{\CP\pm}K^-) + \Gamma(B^+{\to}\Dz_{\CP\pm}K^+)}
			    {\left[\Gamma(B^-{\to}\Dz K^-)+\Gamma(B^+{\to}\Dzb K^+)\right]/2}\,,\ \ \\ 
	\label{eq:Acp}
	A_{\CP\pm} &=& \frac{\Gamma(B^-{\to}\Dz_{\CP\pm}K^-)-\Gamma(B^+{\to}\Dz_{\CP\pm}K^+)}
			    {\Gamma(B^-{\to}\Dz_{\CP\pm}K^-)+\Gamma(B^+{\to}\Dz_{\CP\pm}K^+)}\,.\ \ 
\end{eqnarray}

Here, $\Dz_{\CP\pm} = (\Dz\pm\Dzb)/\sqrt{2}$ are the \CP eigenstates of
the neutral $D$ meson system, following the notation in
Ref.~\cite{gronau1998}. Neglecting $\Dz{-}\Dzb$ mixing~\cite{dmixing},
the observables $R_{\CP\pm}$ and $A_{\CP\pm}$ are related to the angle
$\gamma$, the magnitude ratio $r$ of the amplitudes for the processes $B^-{\to} \Dzb K^-$ and
$B^-{\to} \Dz K^-$, and the relative strong phase
$\delta$ of these amplitudes through the relations $R_{\CP\pm}=1+r^2\pm
2r\cos\delta\cos\gamma$ and $A_{\CP\pm}=\pm
2r\sin\delta\sin\gamma/R_{\CP\pm}$~\cite{glw1991}. Theoretical
predictions for $r$ are on the order of 0.1~\cite{glw1991}, in agreement
with recent results by \babar\ ($r=0.091 \pm 0.059$~\cite{rB_babar})
and Belle ($r=0.159 \pm 0.074$~\cite{belle_dalitz}), obtained through
the study of \btodk, $\Dz\to K^+\pi^-\pi^0$ and $\Dz\to\KS\pi^+\pi^-$
decays.

This analysis, based on $348\,\invfb$ of data collected at the \Y4S
resonance, updates a previous \babar\ study based on $211\,\invfb$ of
data~\cite{babar_dkglw2006}. Belle recently presented a similar
measurement of $R_{\CP\pm}$ and $A_{\CP\pm}$ based on $251\,\invfb$ of
data~\cite{belle_dkglw2006}.

The ratios $R_{\CP\pm}$ are computed under the assumption
$R_{\CP\pm}=R_{\pm}/R$, which holds neglecting a factor of
$r_\pi \lesssim 0.012$ as discussed later. The quantities $R_+$, $R_-$, and $R$
are defined as:
\begin{equation}\label{eq:ratios}
	R_{(\pm)}=\frac{\BR(B^-{\to}\Dz_{(\CP\pm)}K^-)+\BR(B^+{\to}\Dzb_{(\CP\pm)}K^+)}
		       {\BR(B^-{\to}\Dz_{(\CP\pm)}\pi^-)+\BR(B^+{\to}\Dzb_{(\CP\pm)}\pi^+)}\,.
\end{equation}
Several systematic uncertainties affect the $\Dz K$ and $\Dz \pi$ final
states in the same way and therefore cancel in the double ratios $R_{\CPp}$ and $R_{\CPm}$, for
instance the uncertainties on charged particle reconstruction efficiencies, and
the uncertainties on the secondary branching ratios of the \Dz decays.
We express the
$\CP$-sensitive observables in terms of three independent quantities $x_+$, $x_-$, and $r$:
\begin{eqnarray}
	x_\pm &=& \frac{R_{\CPp}(1\mp A_{\CPp})-R_{\CPm}(1\mp A_{\CPm})}{4}\,,\label{eq:x}\\
	r^2 &=& x_\pm^2+y_\pm^2=\frac{R_{\CPp}+R_{\CPm}-2}{2}\,,\label{eq:r2}
\end{eqnarray}
where $x_\pm=r\cos(\delta\pm\gamma)$ and
$y_\pm=r\sin(\delta\pm\gamma)$ are the so called Cartesian
coordinates related to the \CP parameters that are measured using a
Dalitz analysis of \btodk, $\Dz\to\KS\pi^-\pi^+$ decays~\cite{babar_dalitz, belle_dalitz}. This
choice allows the results of the two measurements to be expressed in a
consistent manner.

The measurements use a sample of 382 million \FourS\ decays into
$B\overline{B}$ pairs collected with the \babar\
detector~\cite{detector} at the \pep2 asymmetric-energy $B$ factory.
Charged-particle tracking is provided by a five-layer double-sided
silicon vertex tracker and a 40-layer drift chamber (DCH). A
ring-imaging Cherenkov detector (DIRC) provides additional particle
identification (PID). Photons are identified by the electromagnetic
calorimeter (EMC), which is comprised of 6580 thallium-doped CsI crystals.
These systems are mounted inside a $1.5\,\textrm{T}$ solenoidal superconducting
magnet. We use the GEANT~\cite{geant} software to simulate interactions
of particles traversing the detector, taking into account the varying
accelerator and detector conditions. 

We reconstruct \btodh\ decays, where the prompt track $h^-$ is either a
kaon or a pion. The \Dz candidates are reconstructed in the \CP-even
eigenstates $\pi^-\pi^+$ and $K^-K^+$ ($\Dz_{\CPp}$), in the \CP-odd
eigenstates $\KS\pi^0$ and $\KS\omega$ ($\Dz_{\CPm}$), and in the
(non-\CP) flavor eigenstate $K^-\pi^+$. The $\omega$ candidates are
reconstructed in the $\pi^-\pi^+\pi^0$ channel, and \KS candidates
in the $\pi^+\pi^-$ channel. Compared to the previous
analysis~\cite{babar_dkglw2006}, the current study does not include the
decay mode $\Dz\to\KS\phi$, since it is going to be
explored by a \babar\ Dalitz analysis of $B^- \to \Dz K^-$, $\Dz
\to \KS K^+ K^-$ decays. Excluding the $\KS\phi$ channel from the present analysis will
allow the results of both studies to be more easily combined in the
future.

We optimize our event selection to minimize the statistical error on the
$B^-{\to} \Dz K^-$ signal yield, determined for each \Dz decay channel
using simulated signal and background events. We
reject a candidate track if its Cherenkov angle does not agree within
four standard deviations ($\sigma$) with either the pion or kaon
hypothesis~\cite{PDG2006}, or if it is identified as an electron by the DCH and the
EMC. Neutral pions are reconstructed by combining pairs of photon
candidates with energy deposits larger than 30~\mev that are not
matched to charged tracks. The photon pair invariant mass is
required to be in the range 115--150~\mevcc and the total \piz energy
must be greater than 200~\mev in the laboratory frame. To improve
momentum resolution, the invariant mass of the two photons from
candidate $\piz$'s is constrained to the nominal \piz
mass~\cite{PDG2006}. Neutral kaons are reconstructed from pairs of
oppositely charged tracks with invariant mass within $7.8\,\mevcc$ ($\sim
3\,\sigma$) of the nominal \KS mass. The ratio between the candidate \KS flight
length and its uncertainty must be greater than 2.
The $\omega$ mesons are
reconstructed from $\pi^+\pi^-\pi^0$ combinations with invariant mass
in the range $0.763<M(\pi^+\pi^-\pi^0)<0.799\,\gevcc$. We define
$\theta_N$ as the angle between the normal to the $\omega$ decay plane
and the $\Dz$ momentum in the $\omega$ rest frame, and
$\theta_{\pi\pi}$ as the angle between the flight direction of one of
the three pions in the $\omega$ rest frame and the flight direction of
one of the other two pions in the two-pion rest frame. The
quantities $\cos\theta_N$ and $\cos\theta_{\pi\pi}$ follow
$\cos^2\theta_N$ and $\sin^2\theta_{\pi\pi}$ distributions for the
signal and are almost flat for wrongly reconstructed or false $\omega$
candidates. We require the product
$\cos^2\theta_N\sin^2\theta_{\pi\pi}>0.08 $. The invariant mass of a
\Dz\ candidate $M(\Dz)$ must be within $2.5\,\sigma$ of the mean
fitted mass, with $\sigma$ ranging from $4$ to $20\,\mevcc$ depending on
the $\Dz$ decay mode. To improve the \Dz\ momentum resolution, the
candidate invariant mass is then constrained to the nominal \Dz\
mass~\cite{PDG2006} for all \Dz\ decay channels. For \dotopp, the
invariant mass of the $(h^-\pi^+)$ system, where $\pi^+$ is the pion
from the $\Dz$ and $h^-$ is the prompt track from $B^-$ taken with the
kaon mass hypothesis~\cite{PDG2006}, must be greater than $1.9\,\gevcc$ to reject
background from $B^-{\to}\Dz\pi^-$, $\Dz{\to}K^-\pi^+$ and
$B^-{\to}K^{*0}\pi^-$, $K^{*0}{\to}K^-\pi^+$ decays. We reconstruct $B$
meson candidates by combining a \Dz\ candidate with a track $h$. For
the $\Dz{\to}K^-\pi^+$ mode, the charge of the track $h$ must match
that of the kaon from the $\Dz$ meson decay, selecting $\b\to\c$ mediated $B$ decays.

We select $B$ meson candidates using the
energy difference $\Delta E=E^*_B - E_{ee}^*/2$
and the
beam-energy-substituted mass
$\mes = \sqrt{(E_{ee}^{*2}/2 +
\mathbf{p}_{ee}\cdot\mathbf{p}_B)^2/E_{ee}^2-p_B^2}$,
where the
subscripts $ee$ and $B$ refer to the initial \epem\ system and the $B$
candidate, respectively, and the asterisk denotes the $\epem$
center-of-mass (CM) frame. The \mes distributions for \btodh\ signals
are Gaussian functions centered at the $B$ mass with a resolution of
$2.6\,\mevcc$, and do not depend on the \Dz decay mode or on the nature
of the prompt track. In contrast, the \DeltaE\ distributions depend on
the mass assigned to the prompt track. We evaluate $\Delta E$ with the
kaon mass hypothesis so that the peaks of the distributions are
centered near zero for \btodk\ events and shifted by approximately
$50\,\mev$ for \btodp\ events. The \DeltaE resolution depends on the
momentum resolutions of the \Dz meson and the prompt track $h^-$, and
is typically $16\,\mev$ for all \Dz decay modes under study. All $B$
candidates are selected with \mes within $2.5\,\sigma$ of the mean
value and with \DeltaE in the range $-0.15<\DeltaE<0.20\,\gev$.

To reduce background from $e^+e^- \to q\bar{q}$ events (with $q=u,d,s,c$), 
denoted $q\bar{q}$ in the following, we
construct a linear Fisher discriminant~\cite{fisher} based on the
four event-shape quantities $L_2^{\rm ROE}$, $|\cos\theta_T^*|$, $|\cos\theta_B^*|$
and $R_2^{\rm ROE}$.
The ratio $L_2^{\rm ROE}$ between $L_2=\sum_i p_i\cos^2\theta_i$ and $L_0=\sum_i p_i$ is
evaluated in the CM frame, where the ${\bf p}_i$ are the momenta of charged
tracks and neutral clusters not used to reconstruct the $B$ (i.e., the
rest of the event, ROE), and the $\theta_i$ are their angles with respect to
the thrust axis of the $B$ candidate's decay products.
The angle $\theta_T^*$ is measured between the thrust axis of
the $B$ candidate's decay products and the beam axis, and is evaluated in the CM
frame.
The angle $\theta_B^*$ is measured between the $B$ candidate
momentum and the beam axis, again evaluated in the CM frame.
The ratio $R_2^{\rm ROE}$ of the Fox-Wolfram moments $H_2$ and $H_0$,
is computed using tracks and photons in the ROE~\cite{fox_wol}.
The efficiency of the requirement on the value of the Fisher discriminant
ranges from 74\% to 78\% for \btodk signal events and from 17\% to
23\% for \qqbar background events. For the $K\pi$ channel, the values
are 87\% for signal and 42\% for background events. 

For events with multiple \btodh candidates (0.4\%--7.7\% of the selected
events, depending on the \Dz decay mode), we choose the $B$ candidate with
the smallest $\chi^2 = \sum_c
(M_c-\left<M_c\right>)^2/(\sigma^2_{M_c}+\Gamma^2_{c})$ formed from the
measured and true masses of the composite candidates $c$, $M_c$ and
$\left<M_c\right>$, scaled by the resolution $\sigma_{M_c}$ and width
$\Gamma_{c}$ of the reconstructed mass distributions. Composite candidates considered are
the $B$ candidate itself ($\mes$), $\Dz$, $\pi^0$, and $\omega$ candidates. Also $\Gamma_{\omega}$ is the only
non-negligible width.

The total reconstruction efficiencies, based on
simulated $B\to \Dz K$ events, are 
36\% ($K^-\pi^+$), 
29\% ($K^-K^+$), 
29\% ($\pi^-\pi^+$), 
15\% ($\KS\pi^0$), and 
6\% ($\KS\omega$).  

The main contributions to the background from \BB events come from the
processes $B^-{\to}D^{*}h^-$, $B^-{\to}\Dz\rho^-$, misreconstructed
\btodh, and from charmless $B$ decays to the same final state as the
signal: for instance, the process $B^-{\to}K^-K^+K^-$ is a background
for $B^-{\to}\Dz K^-$, $\Dz{\to}K^-K^+$. These charmless backgrounds
have similar \DeltaE\ and \mes distributions as the $\Dz K^-$ signal and
are referred to in the following as peaking \BB backgrounds ($B^- \to
X_1 X_2 K^-$).

We determine the signal and background yields for each \Dz decay mode
independently from a two-dimensional extended unbinned
maximum-likelihood fit to the selected data events. The fit is
performed simultaneously on the $B^+$ and $B^-$ subsamples. The input
variables to the fit are \DeltaE and the Cherenkov angle $\theta_C$ of
the prompt track as measured by the DIRC. The extended likelihood
$\mathcal{L}$ for $N$ candidates is given by the product of the probabilities for
each individual candidate $i$ and a Poisson factor:
\begin{equation}\label{eq:ext_like}
	\mathcal{L}=\frac{e^{-N'}(N')^N}{N!}\prod_{i=1}^{N}\mathcal{P}_i(\Delta E,\theta_C).
\end{equation}
The probability $\mathcal{P}_i$ is the
sum of the signal and background terms,
\begin{equation}\label{eq:pdf}
	\mathcal{P}_i(\Delta E,\theta_C) = \sum_{J} \frac{N_{J}}{N'}\ \mathcal{P}^J_{\Delta E,i}\ \mathcal{P}^J_{\theta_C,i}\ \ ,
\end{equation}
where $J$ denotes the seven signal and background hypotheses $\Dz h$,
$q\bar{q}(h)$, $B\bar{B}(h)$, and $X_1X_2K$. $N'$ is the total event
yield estimated by the fit, and $N_J$ is the event yield in each
category. We fit directly for the ratios $R' \equiv R_{(\pm)}$ and asymmetries
$A_{\CP\pm}$, as appropriate to the decay mode; they enter
Eq.~(\ref{eq:pdf}) through
\begin{eqnarray}
	N_{\Dz\pi^\pm} & = & \frac{1}{2} \left(1 \mp A_{\CP}^{\Dz\pi}\right) N_{\Dz\pi} \ ,\\
	N_{\Dz K^\pm} & = & \frac{1}{2} \left(1 \mp A_{\CP}\right) N_{\Dz\pi} R' \ ,
\end{eqnarray}
where $N_{\Dz\pi} = N_{\Dz\pi^+} + N_{\Dz\pi^-}$ and $A_{\CP\pm}^{\Dz\pi}$ is defined analogously to Eq.~\ref{eq:Acp}.

The \DeltaE\ distribution for \btdk\ signal is
parameterized with a double Gaussian function. The fraction of the wide component of the signal shape, its offset from
the narrow component and the ratio between the widths of the two
components are fixed to values obtained from simulation. The \DeltaE
probability density function (PDF) for \btdp\ is the same as the \btdk\ one, but
with an additional shift, $\DeltaE_{\rm shift}$, which arises from the
wrong mass assignment to the prompt track. The shift is computed event
by event as a function of the prompt track momentum $p$ and a
Lorentz factor $\gamma_{\textrm{PEP-II}}=E_{ee}/E_{ee}^*$ characterizing the boost to the $\epem$ CM frame:
\begin{equation}
	\DeltaE_{\rm shift} = \gamma_{\textrm{PEP-II}} \left(\sqrt{m_K^2+p^2} -
	\sqrt{m_{\pi}^2+p^2}\right).
\end{equation}
The
\DeltaE\ distributions for the continuum background are parameterized
with a straight line. The \DeltaE\ distribution for the \BB
background is empirically parametrized with a Gaussian peak with an exponential
tail~\cite{cb}. The parameters of the background shapes are determined from
simulated events (\BB) and off-resonance data (\qqbar) and are fixed in
the fit. The number of peaking background events $N_{X_1X_2 K}$ is
fixed to values obtained from a study of the $\Dz$ mass sidebands. The
particle identification PDF is a double Gaussian as a function of
$\theta_C^{\rm pull}$, which is the difference between the measured
Cherenkov angle $\theta_C$ and its expected value for a given mass
hypothesis, divided by the estimated error. The PID shape parameters are
obtained from simulation.
To summarize, the floating parameters in each of the five
the fits are the $\Dz K$ and $\Dz \pi$ signal yield asymmetries, the
total number of signal events in $\Dz \pi$, the appropriate ratios $R$ and $R_{\pm}$,
eight background yields (one for each charge), and two parameters of
the \DeltaE signal shape (common for positive and negative
samples).

The results of the fits, expressed in terms of signal yields, are
summarized in Table~\ref{tab:fitresults}. Figure~\ref{fig:fit_kaons}
shows the distributions of \DeltaE\ for the $K^-\pi^+$, \CPp\ and \CPm\
modes after enhancing the \btodk purity by requiring that the
prompt track be consistent with the kaon hypothesis. This requirement
is $88\%$ ($1\%$) efficient for $h^-=K^-$ ($h^-=\pi^-$).

\begin{table}[h]
\caption{Uncorrected yields as obtained from the maximum likelihood fit. The quoted uncertainties are statistical.}
\label{tab:fitresults}
\begin{center}
\begin{tabular}{lccccc}
\hline
\hline
$\Dz$     & \CP & $N(D\pi^+)$ & $N(D\pi^-)$ & $N(DK^+)$ & $N(DK^-)$\\
\hline
$K^-\pi^+$    &     & $   12745\pm120     $ & $   12338\pm120     $ & $     954\pm36      $ & $     918\pm36      $\\ 
\hline
$K^-K^+$      & $+$ & $    1109\pm36      $ & $    1051\pm35      $ & $      51\pm10      $ & $     113\pm13      $\\
$\pi^-\pi^+$  & $+$ & $     390\pm24      $ & $     378\pm24      $ & $      39\pm9       $ & $      36\pm9       $\\
\hline
$\KS\pi^0$    & $-$ & $    1102\pm37      $ & $    1134\pm38      $ & $     100\pm13      $ & $      88\pm12      $\\ 
$\KS\omega$   & $-$ & $     422\pm24      $ & $     403\pm26      $ & $      29\pm8       $ & $      18\pm8       $\\
\hline
\hline
\end{tabular}
\end{center}
\end{table}

\begin{figure}[!htb]
\begin{center}
\includegraphics[width=7.5cm]{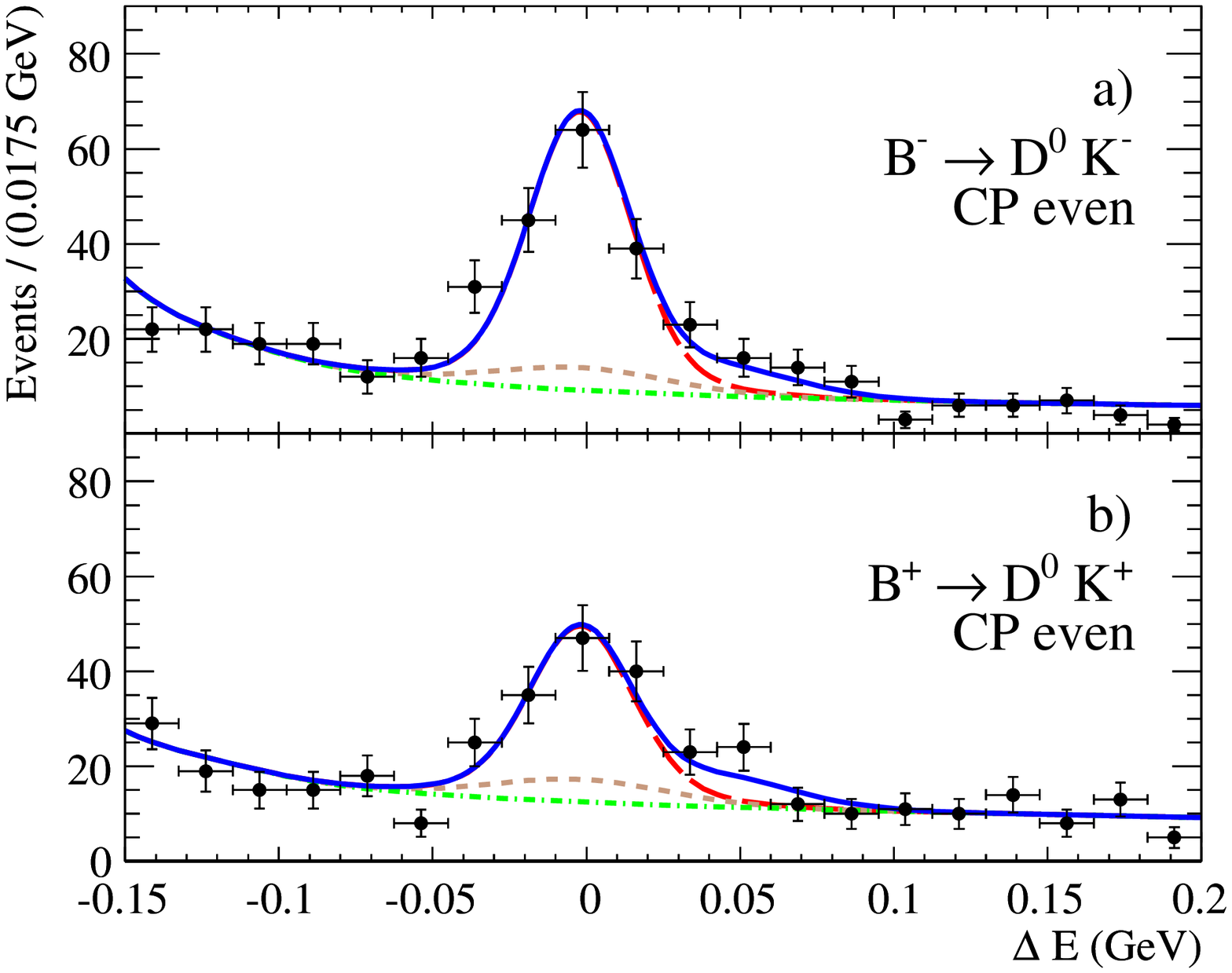}
\includegraphics[width=7.5cm]{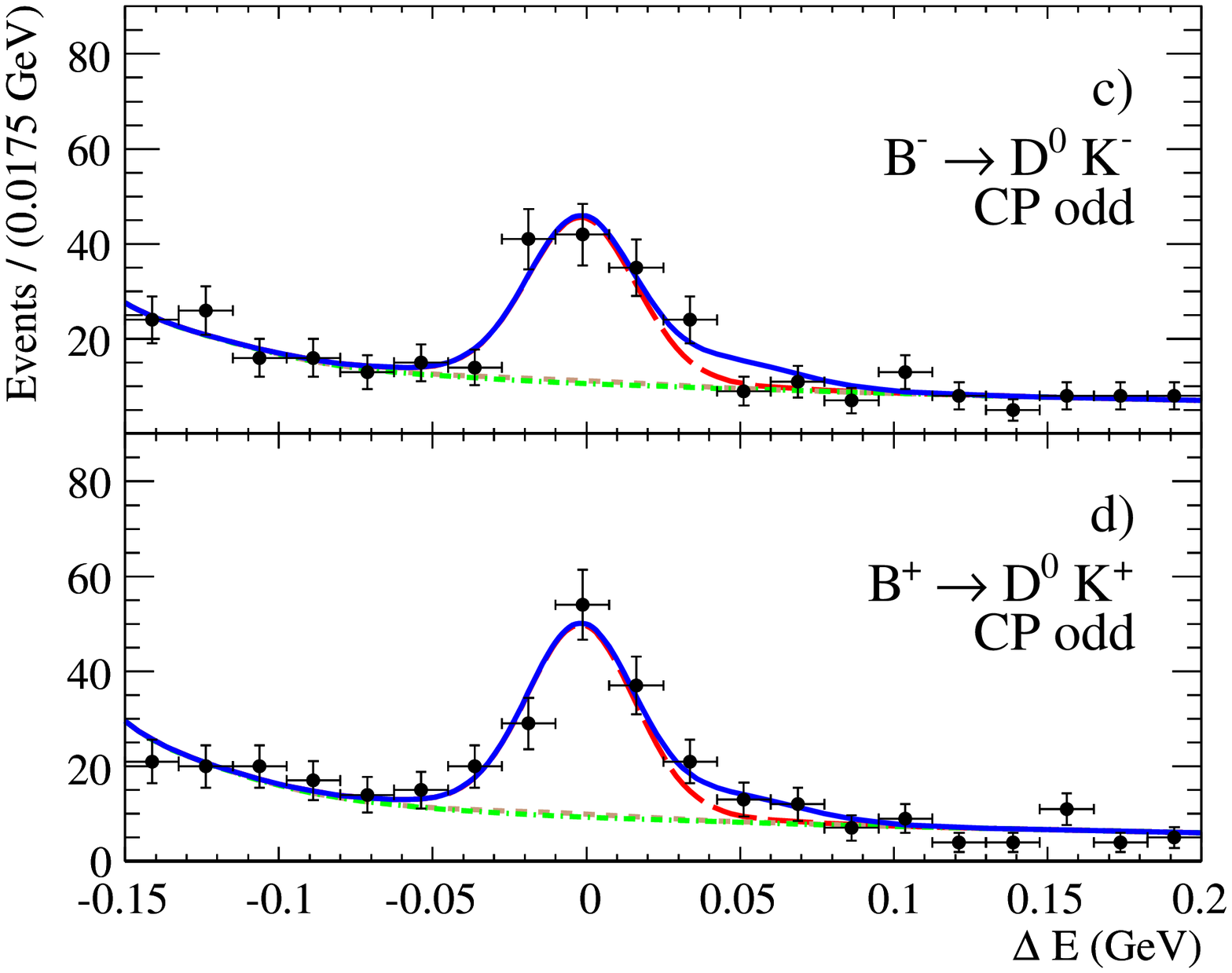}
\includegraphics[width=7.5cm]{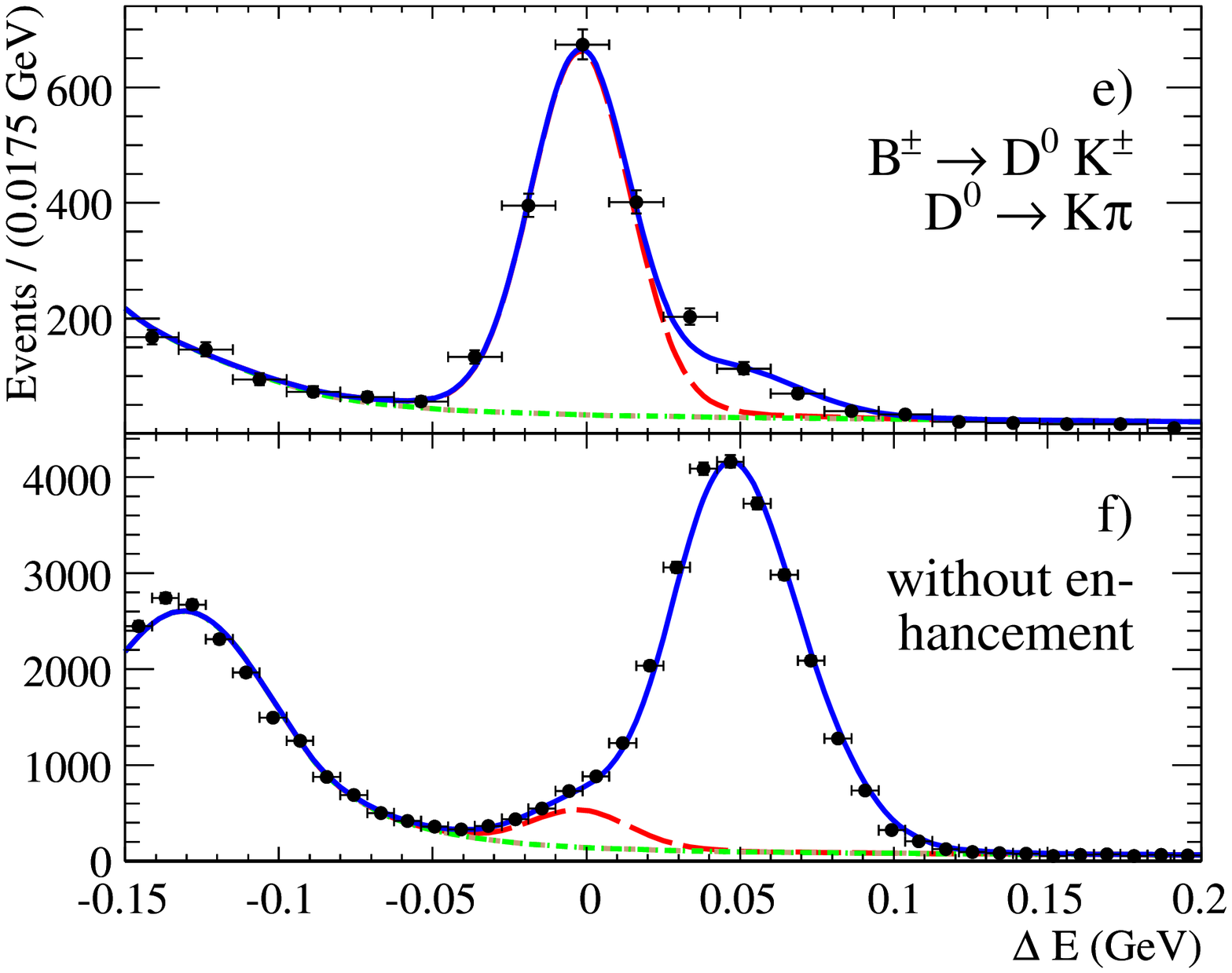}
\caption{Distributions of \DeltaE\ for events enhanced in
  $B^\pm\to\Dz K^\pm$ signal:
  a) $B^-\to \Dz_{\CPp}K^-$;
  b) $B^+\to \Dz_{\CPp}K^+$;
  c) $B^-\to \Dz_{\CPm}K^-$;
  d) $B^+\to \Dz_{\CPm}K^+$;
  $B^\pm\to\Dz K^\pm$, $\Dz{\to}K^\pm\pi^\mp$ with (e) and without (f) signal enhancement.
  Blue (continuous) curve: projection of the full PDF of the maximum likelihood fit.
  Red (long-dashed): $B^\pm\to\Dz K^\pm$ signal on all backgrounds.
  Brown (short-dashed): peaking component on $q\bar{q}$ and $B\bar{B}$ background.
  Green (dash-dotted): $q\bar{q}$ and $B\bar{B}$ background.
  }
\label{fig:fit_kaons}
\end{center}
\end{figure}

The ratios $R_{(\pm)}$, as measured by each fit, are corrected to take
into account small differences in the selection efficiency between
$B{\to}DK$ and $B{\to}D\pi$. The efficiency ratios range from
$1.013\pm0.006$ to $1.037\pm0.010$. Their uncertainties are due to the
statistics of the simulated samples and are considered in the study of
systematic uncertainties.
In the case of $\Dz\to\KS \omega$, $\omega{\to}\pi^+\pi^-\pi^0$, the values
of $R_{\CP-}^{\KS\omega}$ and $A_{\CP-}^{\KS\omega}$ need to be
corrected to take into account a possible dilution from a non-resonant
\CP-even background arising from $B^-{\to}\Dz h^-$,
$\Dz\to\KS(\pi^-\pi^+\pi^0)_{\textrm{non}-\omega}$ decays. There is
little information on this background. We estimate the corrections
using a fit to the $\omega$ helicity angle in the selected data events
and find the correction factors to be $1.12 \pm 0.14$ for
$A_{\CP-}^{\KS\omega}$ and $1.00 \pm 0.01$ for $R_{\CP-}^{\KS\omega}$.
The uncertainties in the correction factors are included in the
systematic errors.
After applying all corrections, the quantities $R_{\pm}/R$ and
$A_{\CP\pm}$ are computed by means of a weighted average over the
$\CPp$ and $\CPm$ modes. The results for the \CP-even and \CP-odd
combinations are reported in Table~\ref{tab:final_ratio}.

Systematic uncertainties in $R_{\CP\pm}$ and $A_{\CP\pm}$ are listed in Table~\ref{tab:syst}. The
uncertainties on the fitted signal yields are due to the imperfect
knowledge of the $\DeltaE$ and PID PDFs and of the peaking background
yields, and are evaluated in test fits by varying the parameters of the
PDFs and the peaking background yields by $\pm1\sigma$ and taking the
difference in the fit results. A possible $\pm20\%$ \CP asymmetry in
the peaking background is considered in the same way. In the
$\KS\omega$ channel we also take into account the uncertainties in
the correction factors due to the \CP-even backgrounds from
$\Dz\to\KS(\pi^-\pi^+\pi^0)_{\textrm{non}-\omega}$ decays. A
possible bias in the measured $A_{\CP\pm}$ comes from an intrinsic
detector charge asymmetry due to asymmetries in acceptance or tracking
and particle identification efficiencies. An upper limit on this bias
is obtained from the measured asymmetries in the processes
$B^-{\to}\Dz h^-$, $\Dz{\to}K^-\pi^+$ and $B^-{\to}\Dz_{\CP\pm}\pi^-$,
where \CP violation is expected to be negligible. From the average
asymmetry, $-(1.6\pm 0.6)\%$, we obtain the limit $\pm 2.2\%$ for the
bias. For the branching fraction ratios $R_{\CP\pm}$, an additional
source of uncertainty is associated with the assumption that $R_{\CP\pm} = R_{\pm}/R$.
This assumption holds only if the magnitude of the ratio $r_\pi$ between
the amplitudes of the $B^-{\to} \Dzb\pi^-$ and $B^-{\to} \Dz\pi^-$ processes is neglected~\cite{gronau2003}. $r_\pi$ is expected to be small:
$r_\pi \sim r \frac{\lambda^2}{1-\lambda^2} \lesssim 0.012$, where
$\lambda \approx 0.22$~\cite{PDG2006} is the sine of the Cabibbo angle.
This introduces a relative uncertainty $\pm 2 r_\pi \cos\delta_\pi
\cos\gamma$ on $R_{\CP\pm}$, where $\delta_\pi$ is the relative strong
phase between the amplitudes $\mathcal{A}(B^-{\to}\Dzb \pi^-)$ and $\mathcal{A}(B^-{\to}\Dz
\pi^-)$. Since $|\cos\delta_\pi \cos\gamma|\le 1$ and $r_\pi \lesssim
0.012$, we assign a relative uncertainty $\pm 2.4\%$ to $R_{\CP\pm}$,
which is completely anti-correlated between $R_{\CPp}$ and $R_{\CPm}$.
We quote the measurements in terms of $x_\pm$ and $r^2$,
\begin{eqnarray}
	x_+ & = & -0.09 \pm 0.05\stat \pm 0.02\syst\,,\\
	x_- & = & +0.10 \pm 0.05\stat \pm 0.03\syst\,,\\
	r^2 & = & +0.05 \pm 0.07\stat \pm 0.03\syst.
\end{eqnarray}
The correlations between the different sources of systematic errors,
when non-negligible, are considered when calculating $x_\pm$ and $r^2$.
The measured values of $x_\pm$ are consistent with those found from
\btodk, $\Dz\to\KS\pi^-\pi^+$ decays, and the precision is
comparable~\cite{babar_dalitz}.

\begin{table}[h]
\caption{Measured ratios $R_{\CP\pm}$ and $A_{\CP\pm}$
for $\CP$-even (\CPp) and $\CP$-odd (\CPm) $D$ decay modes.
The first error is statistical; the second is
systematic.}
\label{tab:final_ratio}
\begin{center}
\begin{tabular}{lcc}
\hline
\hline
$\Dz$ mode \ \ & $R_{\CP}$ & \ \ $A_{\CP}$\\
\hline
$\CPp$ \ \ & $1.06\pm 0.10\pm 0.05$ & \ \ \ \ $0.27\pm 0.09\pm 0.04$ \\
$\CPm$ \ \ & $1.03\pm 0.10\pm 0.05$ & \ \    $-0.09\pm 0.09\pm 0.02$ \\
\hline
\hline
\end{tabular}
\end{center}
\end{table}

\begin{table}[h]
\caption{Systematic uncertainties on the observables $R_{\CP\pm}$ and
  $A_{\CP\pm}$ in absolute terms.}
\label{tab:syst}
\begin{center}
\begin{tabular}{lcccc}
\hline
\hline
source                            & $\Delta R_{\CPp}$ & $\Delta R_{\CPm}$ & $\Delta A_{\CPp}$ & $\Delta A_{\CPm}$ \\
\hline
fixed fit parameters              & $0.036$	      & $0.019$           & $0.010$	      & $0.002$ 	  \\
peaking background                & $0.029$	      & $0.037$           & $0.031$	      & $0.003$ 	  \\
detector charge asym.             & -		      & -	          & $0.022$	      & $0.022$ 	  \\
opp. \CP bkg. in $\KS\omega$      & -		      & $0.002$           & -		      & $0.007$ 	  \\
$R_{\CP\pm}$ vs. $R_\pm$          & $0.026$	      & $0.025$           & -		      & -		  \\
$K/\pi$ efficiency                & $0.002$           & $0.007$           & -                 & -                 \\
\hline
total                             & $0.053$	      & $0.049$           & $0.039$	      & $0.023$ 	  \\
\hline
\hline
\end{tabular}
\end{center}

\end{table}

In conclusion, we have reconstructed \btodk\ decays with \Dz mesons
decaying to non-\CP, \CP-even and \CP-odd eigenstates.
The combined uncertainties we find for $A_{\CP\pm}$ ($R_{\CP\pm}$)
are smaller by a factor of $0.7$ ($0.9$) and $0.6$ ($0.6$) than the previous
\babar~\cite{babar_dkglw2006} and Belle~\cite{belle_dkglw2006} measurements, respectively.
We find $A_{\CPp}$ to deviate by $2.8$ standard deviations from zero.
We express the results in terms of the
Cartesian coordinates $x_\pm$ and $r^2$ (Eqs.~\ref{eq:x},
\ref{eq:r2}).
These measurements, combined with the
existing measurements from \btodk decays, will improve our knowledge
of the angle $\gamma$ and the parameter $r$.

We are grateful for the excellent luminosity and machine conditions
provided by our \pep2\ colleagues, 
and for the substantial dedicated effort from
the computing organizations that support \babar.
The collaborating institutions wish to thank 
SLAC for its support and kind hospitality. 
This work is supported by
DOE
and NSF (USA),
NSERC (Canada),
CEA and
CNRS-IN2P3
(France),
BMBF and DFG
(Germany),
INFN (Italy),
FOM (The Netherlands),
NFR (Norway),
MES (Russia),
MEC (Spain), and
STFC (United Kingdom). 
Individuals have received support from the
Marie Curie EIF (European Union) and
the A.~P.~Sloan Foundation.


\begin{thebibliography}{99}
\bibitem{chargeconj}
Reference to the charge-conjugate state is implied
here and throughout the text unless otherwise stated.
\end{thebibliography}
\end{document}